\documentclass[twocolumn,prl,superscriptaddress, preprintnumbers,amsmath,amssymb,floatfix]{revtex4}
\pdfoutput=1
\usepackage{graphics}
\usepackage{graphicx}
\usepackage{amssymb}		
\usepackage{amsmath}
\usepackage{verbatim}
\usepackage{pstricks}
\newcommand{\bea}{\begin{eqnarray}}
\newcommand{\eea}{\end{eqnarray}}

\newcommand{\Mev}{{\rm MeV}}
\newcommand{\gev}{{\rm GeV}}

\newcommand{\op}[1]{\ensuremath{{\mathcal O}_{#1}}}

\begin{document}

\preprint{MCTP-09-42}

\title{Momentum Dependent Dark Matter Scattering}

\author{Spencer Chang}
\affiliation{Physics Department,~University of California Davis,
Davis,~California 95616
}

\author{Aaron Pierce}
\affiliation{Michigan Center for Theoretical Physics (MCTP),
Department of Physics, University of Michigan, Ann Arbor, MI
48109
}

\author{Neal Weiner}
\affiliation{Center for Cosmology and Particle Physics,
Department of Physics, New York University, New York, NY
10003
}

\date{\today}

\begin{abstract}
 It is usually assumed that WIMPs interact through spin-independent and spin-dependent interactions. Interactions which carry additional powers of the momentum transfer, $q^2$, are assumed to be too small to be relevant. In theories with new particles at the $\sim \gev$ scale, however, these $q^{2}$-dependent interactions can be large, and, in some cases dominate over the standard interactions. This leads to new phenomenology in direct detection experiments.
Recoil spectra peak at non-zero energies, and the relative strengths of different experiments can be significantly altered.
We present a simple parameterization for models of this type which captures much of the interesting phenomenology and allows a comparison between experiments.  As an application, we find that dark matter with momentum dependent interactions coupling to the spin of the proton can reconcile the DAMA annual modulation result with other experiments.
\end{abstract}

\maketitle
\section{Introduction}
The presence of dark matter (DM) in our universe is now well established by a variety of astrophysical measurements, over a wide range of scales from sub-kpc to Gpc. Its clustering and low interaction cross sections are supported by the success of the CDM framework and the ability of N-body simulations to reproduce observed structures. In spite of these great successes, we remain ignorant to its detailed nature. A direct detection of DM via its recoils off of nuclei would  confirm its particle nature and yield insight into its origin.  An examination of the recoil spectrum would provide important information about its properties, and possibly the formation history of the galaxy. 

If the dark matter is a Majorana fermion $\chi$ - a supersymmetric neutralino being the most prominent example - the types of interactions available are significantly limited.  The dominant scatterings are mediated via the operators:
\begin{eqnarray}
{\mathcal O}_{SI} &= &(\bar{\chi} \chi) (\bar{q} q),  \label{eqn:SI} \\
{\mathcal O}_{SD} &=& (\bar{\chi} \gamma_{\mu} \gamma_{5} \chi) (\bar{q} \gamma^{\mu} \gamma_{5} q) , \label {eqn:SD}
\end{eqnarray}
which give respectively spin-independent (SI) and spin-dependent (SD) scattering.
As these operators typically dominate the interaction rate of Weakly Interacting Massive Particles (WIMPs) with nuclear targets, direct detection experiments quote results as bounds on the spin-independent and spin-dependent cross section per nucleon.

Nonetheless, there are more dimension-6 operators that can contribute to the direct detection cross section. Namely, 
\begin{eqnarray}
{\mathcal O}_{1}&=&(\bar{\chi} \gamma _{5} \chi) (\bar q   q), \label{eqn:operators1} \\
{\mathcal O}_{2}&=&(\bar{\chi}  \chi) (\bar q  \gamma_5 q), \\
{\mathcal O}_{3}&=&(\bar{\chi} \gamma_{5} \chi) ( \bar q  \gamma_{5} q), \\
{\mathcal O}_{4}&=&(\bar{\chi} \gamma_{\mu} \gamma_{5} \chi) (\bar{q} \gamma^{\mu} q).
\label{eqn:operators}
\end{eqnarray}
The operators ${\mathcal O}_{1}$, ${\mathcal O}_{2}$, and ${\mathcal O}_{4}$  are not present if parity is a good symmetry of the theory, but since parity is badly broken in the Standard Model and it could be badly broken in the dark matter sector, it is reasonable to include them.  If $\chi$ is a Dirac fermion, instead of Majorana, additional operators are possible. In particular, there is the possibility of a dipole or charge radius coupling to dark matter and a vector coupling to quarks \cite{cidm, DipolarDM}. Such an operator is quantitatively similar to \op{1}, with the principle difference that it typically couples to atomic number $Z$ rather than mass number $A$. 

These operators in Eqns.~(\ref{eqn:operators1})--(\ref{eqn:operators}) are present even in the context of the minimal supersymmetric Standard Model (MSSM), but there the contributions to scattering are typically far subdominant,
as they are suppressed relative to  $\op{SI/SD}$ by additional powers of momentum $O(q^2/M_W^2) \sim 10^{-6}$, or in the case of ${\mathcal O}_4$ also by velocity suppression $v^2$.  Consequently, they are usually ignored  \cite{FalkFerstl}, but see \cite{Chatto}.   Moreover, even if the dominant operators are zero, because of this suppression, these new operators are typically negligible in the context of direct detection experiments. Thus, even neglecting ${\mathcal O}_{SI/SD}$ it might seem unlikely that such interactions would be relevant for upcoming direct detection experiments. 

However, this reasoning ignores that these two facts often go hand in hand. $\op{SI/SD}$ are typically small when there is a symmetry reason for them to be small, in particular, when the DM-nucleon force is mediated by a pseudo-Goldstone boson (PGB). Because of their shift symmetry, PGBs have $q^2$ suppressed interactions. At the same time, PGBs are also naturally much lighter than the weak scale. Thus, $\op{1-4}$ are no longer insignificant if the mediator has mass $\lesssim O(\gev)$. When we combine this with the recent interest in new GeV-scale particles, e.g., \cite{ArkaniHamed:2008qn}, and in particular PGBs \cite{Nomura:2008ru} arising from models to explain PAMELA, Fermi, ATIC and HESS, we are strongly motivated to consider these scenarios.

In this paper, we explore a class of dark matter models where the scattering is momentum dependent (MDDM), i.e., where the operators \op{1-4} dominate and are large enough to be observable in upcoming direct detection experiments. As we shall see, these operators can have a significant impact on the spectral shape and the sensitivity of various experiments. As an example, we shall see that these effects can improve the ability to explain the DAMA annual modulation signal while being consistent with other direct detection exclusions. 

\section{Signals of MDDM}

The recoil rates at a direct detection experiment can be written as
\begin{eqnarray}
\frac{dR}{dE_R}= \frac{N_T m_N \rho_\chi}{2m_\chi \mu^2} \sigma(q^2) \int^\infty_{v_{min}} \frac{f(v)}{v} dv,
\end{eqnarray}
where $m_N$ is the nucleus mass, $N_T$ is the number of target nuclei in the detector, $\rho_\chi = 0.3$ GeV/cm$^3$ is the WIMP density, $\mu$ is the reduced mass of the WIMP-nuclei system, and $f(v)$ is the halo velocity distribution function in the lab frame.  The minimum velocity to scatter with energy $E_R$ is $v_{min}= \sqrt{m_N E_R/2\mu^2}$.     The rest of the expression depends on the scattering's $q^2 = 2m_N E_R$.  For SI interactions, we have 
\begin{equation}
\sigma(q^2)_{SI} = \frac{4 G_F^2 \mu^2}{\pi}  \left[Z f_p+(A-Z)f_n\right]^2 F^2(q^2),
\end{equation}
where $f_p,f_n$ are respectively the couplings to the proton and neutron, and $F^2$ factor is the form factor.  We take the limit $f_p = f_n$.  This expression is then proportional to the nucleon scattering cross section $\sigma_p = \frac{4}{\pi} G_F^2 \mu_p^2 f_p^2$, where $\mu_p$ is the reduced mass of the WIMP-proton system.   
For SD interactions, we have
\begin{eqnarray}
\sigma(q^2)_{SD}& =& \frac{32 G_F^2 \mu^2}{2J+1} [a_p^2 S_{pp}(q^2)+a_p a_n S_{pn}(q^2)\nonumber \\
& & + a_n^2 S_{nn}(q^2) ],
\end{eqnarray}
where $a_p,a_n$ are respectively the couplings to the proton and neutron, and the $S$ factors are the form factors for SD scattering.  The corresponding nucleon cross sections are $\sigma_{(p,n)} = \frac{24}{\pi}G_F^2 \mu_{(p,n)}^2 a_{(p,n)}^2$.

The effect of the new operators can be parameterized simply:
\begin{equation}
\frac{dR_i^{MDDM}}{dE_R} = \left(\frac{q^{2}}{q^{2}_{ref}}\right)^n  \left(\frac{q^2_{ref}+m^2_{\phi}}{q^2+m^2_{\phi}} \right)^2 \frac{dR_i}{dE_R},\label{eqn:newdR}
\end{equation}
where $i$ indexes the interaction, i.e., SD-proton, SD-neutron or SI, and we have included the propagator due to a light mediator $\phi$ with mass $m_\phi$.   For the benchmark cases, we will take $m_\phi^2 \gg q^2$, to arrive at the simple form
\begin{equation}
\frac{dR_i^{MDDM}}{dE_R} = \left(\frac{q^{2}}{q^{2}_{ref}}\right)^n \frac{dR_{i}}{dE_R}.\label{eqn:newdRsimp}
\end{equation}
 We have chosen to normalize the new factors out front at a reference value $q^2_{ref} \equiv (100\; \text{MeV})^2$,  a characteristic value for many direct detection experiments.  For operators ${\mathcal O}_1, {\mathcal O}_2$ the exponent $n=1$, while for ${\mathcal O}_3$, $n=2$.  For ${\mathcal O}_1$, the interaction is spin-independent on the nucleus side, while for the others it is spin-dependent.  This form of the recoil rate defines the nucleon cross sections $\sigma_{p,n}$ for momentum dependent scattering.    \op{4}'s scattering cannot be written in this form, since it has terms proportional to the DM velocity.  However, we find that its spectra is almost identical to standard SI scattering, so we neglect it for the rest of the paper.

\begin{figure*}[ht]
\begin{center}
a) \includegraphics[width=6cm]{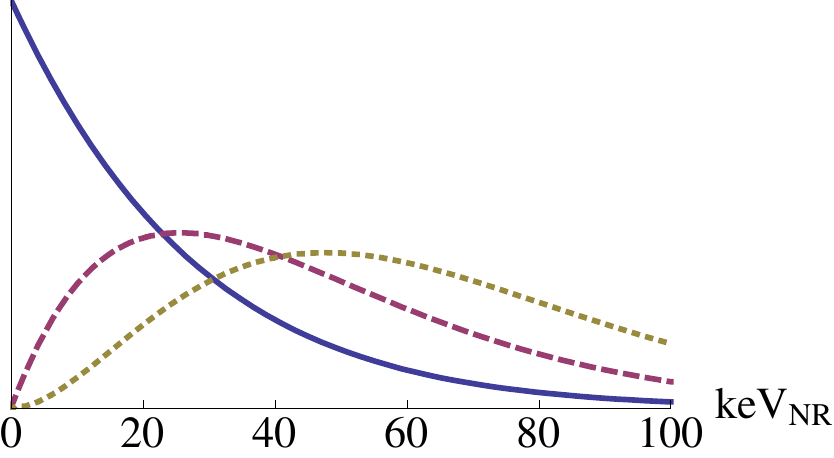} \hspace{1cm }b) \includegraphics[width=6cm]{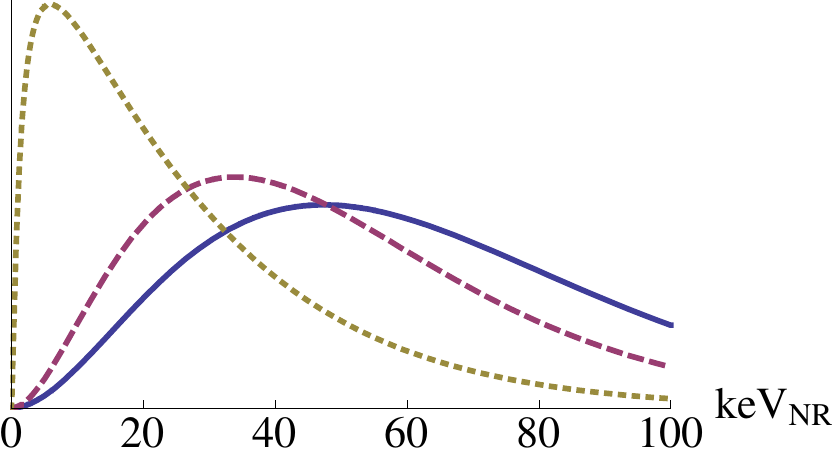}
\end{center}
\caption{Germanium spectra plots with arbitrary normalization versus energy recoil for SI momentum dependent scattering of a 100 GeV dark matter mass.  Plot a) displays the effect of additional powers of $q^2$ with $q^0,q^2,{\rm and\;} q^4$ in solid, long dash and short dash.  Plot b) illustrates the effect of  $m_\phi$ on the $q^4$ suppressed scenario with $m_\phi=(1,.1,.01)\; \gev$ in solid, long dash and short dash. \label{fig:Gespectra} }
\end{figure*}

MDDM is characterized by a modification of its nuclear recoil spectrum. Typically, direct detection experiments optimize their searches by going to lower energy thresholds, where standard WIMP signatures are expected to peak. In contrast, the spectrum of MDDM vanishes at zero recoil energy, and then can be either peaked or fairly flat over the range in question.

We show in Fig.~\ref{fig:Gespectra} the spectra of MDDM scenarios for the case of SI germanium scattering. As we can see, the spectra differ dramatically from those expected for conventional dark matter.  The powers of $q^2$ suppress the low energy events resulting in a peaked spectrum reminiscent of inelastic dark matter (iDM) \cite{TuckerSmith:2001hy,TuckerSmith:2004jv,WeinerKribs}.  In contrast to inelastic dark matter, here the peaking arises without needing a coincidence of parameters (specifically, $\delta$ in iDM models tuned to the WIMP kinetic energy).  The spectrum need not be sharply peaked, however, and can be broadly spread over a large range of recoil energies.  The non-trivial propagator of  Eqn.~(\ref{eqn:newdR}) allows the possibility that events can be suppressed for $q^2\gg m_\phi^2$, as can be seen in Fig.~\ref{fig:Gespectra}.  Finally, increasing the dark matter mass shifts the spectra to higher energies.  Given the possibilities, the lesson is that search strategies developed for the simplest dark matter candidates are by no means optimal for every dark matter candidate.  The true dark matter candidate may not be one of these simplest possibilities, and it is important to cast a wide net.

\section{Existing Searches and DAMA}
\begin{figure*}
a) \includegraphics[width=6cm]{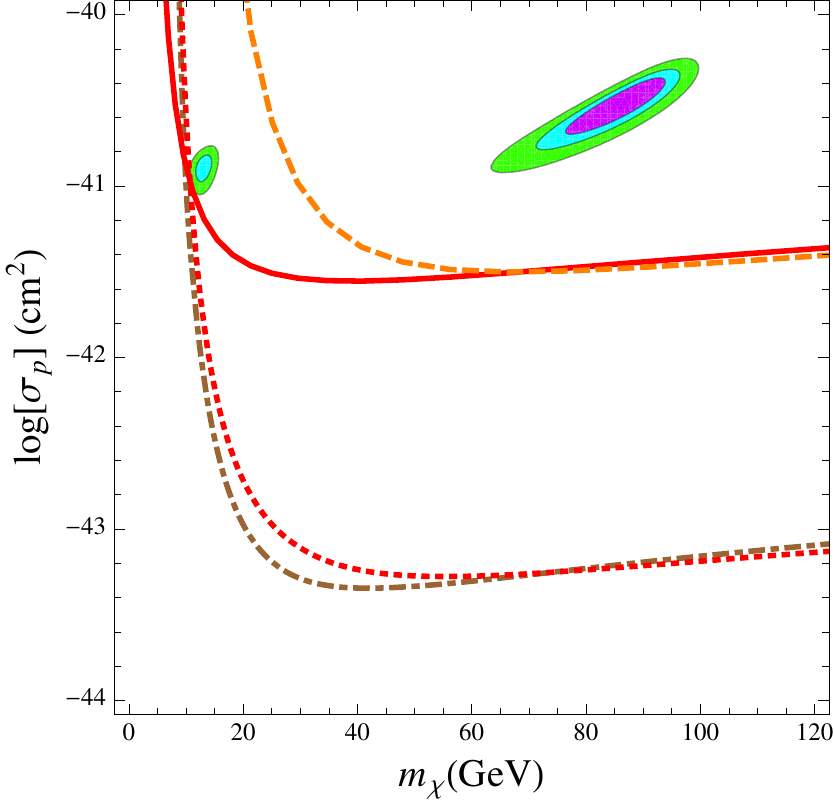}\hspace{1cm}
b) \includegraphics[width=6cm]{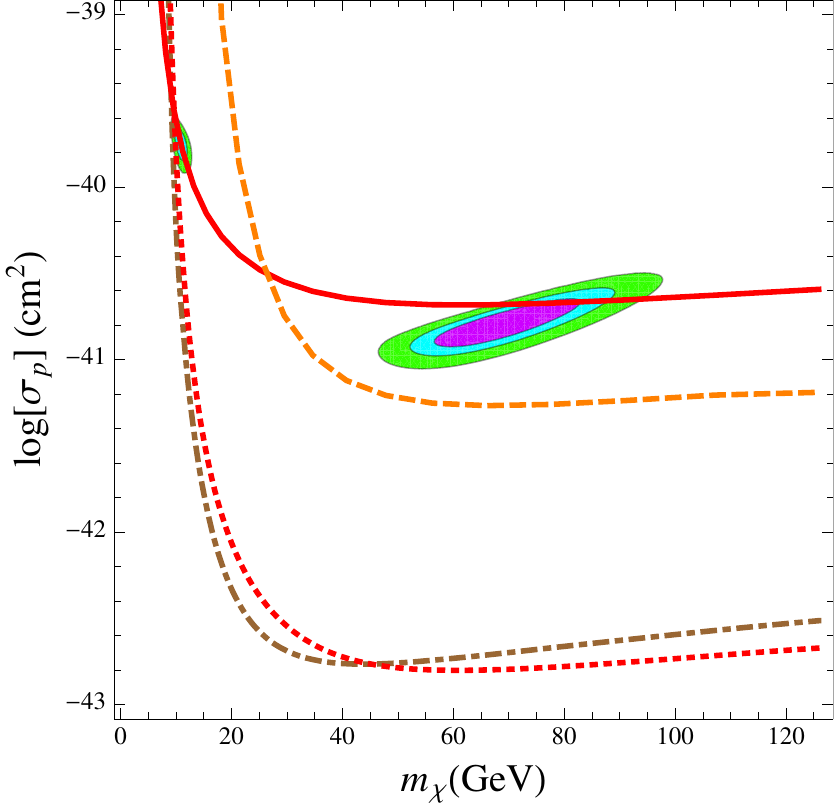}
\caption{Plots of the SI nucleon cross section $\sigma_p$ vs DM mass $m_\chi$ without (a) and with $q^2$ suppression (b).  The colored regions show the 68, 90, and 99\% CL regions for the best DAMA fit. The 90\% exclusions limits are KIMS (orange dashed), CDMS Si (red solid), CDMS Ge (red dotted) and XENON10 (brown dot-dashed). We have taken $f_p=f_{n}$.\label{fig:SIq2} }
\end{figure*}

\begin{figure*}
a) \includegraphics[width=5cm]{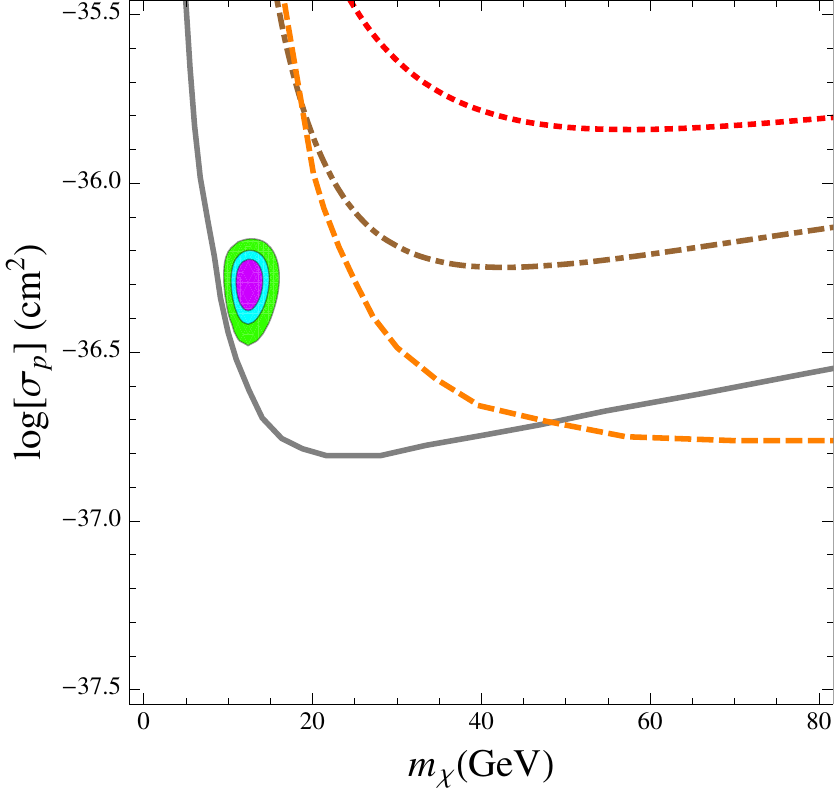}
b) \includegraphics[width=5cm]{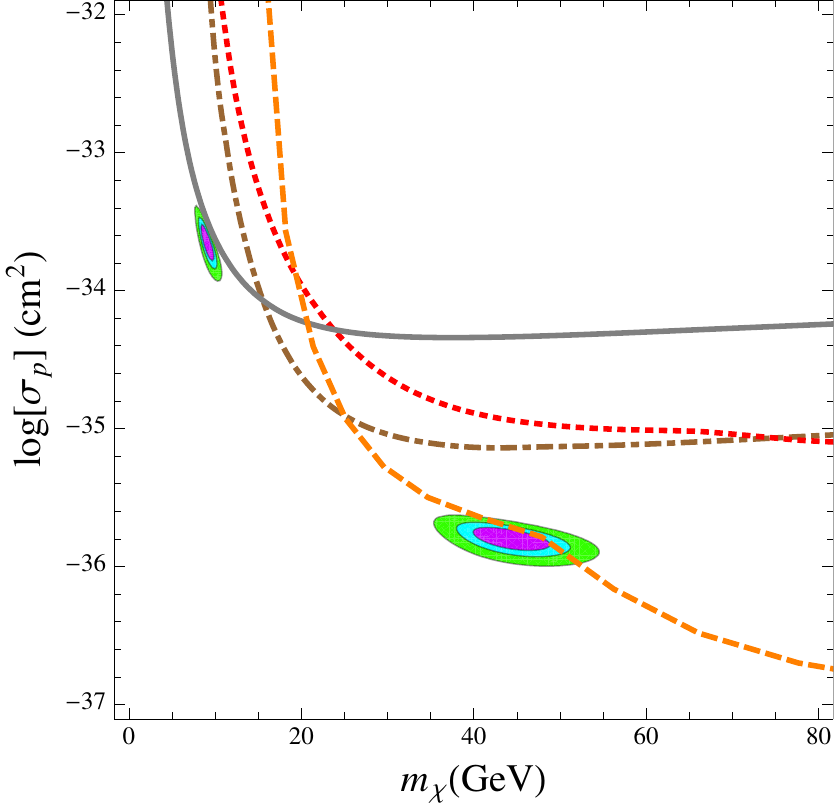}
c) \includegraphics[width=5cm]{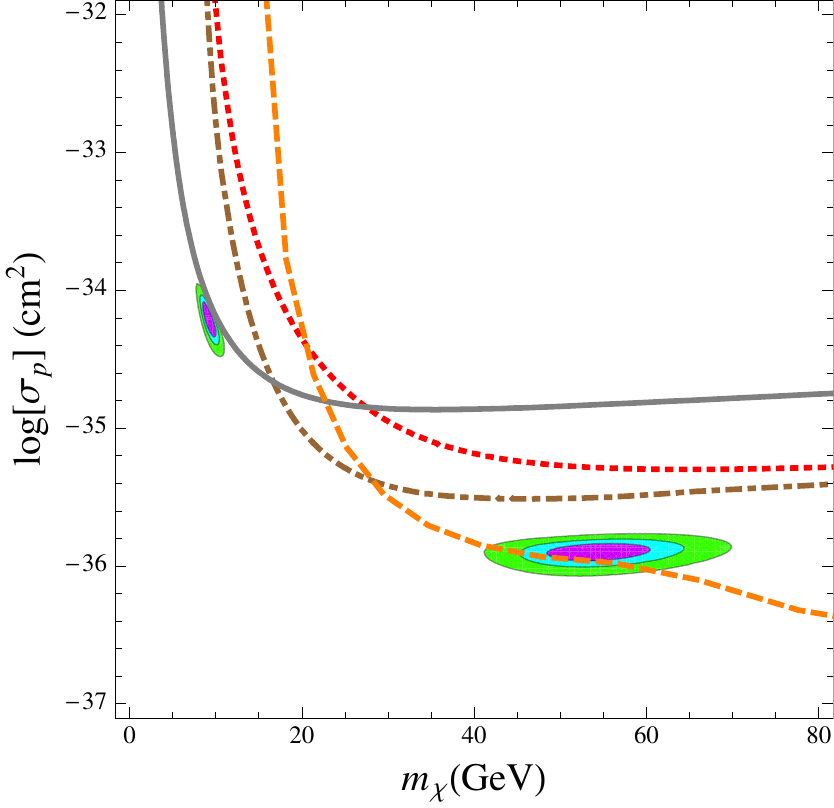}
\caption{Plots of the SD-proton cross section $\sigma_p$ vs DM mass $m_\chi$ without (a) and with $q^4$ suppression (b) and (c), where the mediator mass is 1 GeV (b) and 100 MeV (c).  The colored regions show the 68, 90, and 99\% CL regions for the best DAMA fit. The 90\% exclusions limits are PICASSO (gray solid), KIMS (orange dashed), XENON10 (brown dot-dashed), and CDMS (red dotted). \label{fig:SDp} }
\end{figure*}

To understand the effects of MDDM, we study how these $q^2$ effects can modify the limits arising from existing experiments. We show in Fig.~\ref{fig:SIq2} the limits on interactions mediated by \op{1} compared with limits on standard SI interactions, and in Fig.~\ref{fig:SDp} the limits of \op{3} when compared with standard SD-proton interactions. We follow the procedure laid out in Ref.~\cite{Chang:2008xa}  for CDMS \cite{Akerib:2005kh} and XENON10 \cite{Angle:2007uj} limits and Ref.~\cite{WeinerKribs} for KIMS \cite{KIMS} limits.  Although PICASSO \cite{PICASSO} and COUPP \cite{COUPP} limits are comparable, we only discuss PICASSO limits in what follows.  Our methods better reproduce their (momentum independent) result in the SD-proton case, making us more confident that the MDDM limit is realistic. 

Inspecting the exclusion limits of these plots, we see important changes with respect to the traditional cases. First, consider  the SI case with $q^{2}$ dependence (\op{1}).  While CDMS-Ge and XENON10 remain the strongest, KIMS becomes stronger than CDMS-Si over much of the parameter space. In the SD-proton case, limits from PICASSO are significantly weaker, and XENON10 becomes stronger than KIMS in the 15-25 GeV range.

These results are easy to understand.  Due to suppression of low energy events in the MDDM scenario, experiments that rely upon low energy thresholds (in particular, XENON10) are weakened when compared with others with higher thresholds (such as CDMS), which is why CDMS improves relative to XENON10. 
On the other hand, since $q^2 = 2M_N E_R$, at a given recoil energy, heavier nuclei are preferred by momentum dependent scattering, which is why KIMS improves over CDMS Si and why PICASSO weakens relative to the other experiments.  Another effect occurs for COUPP and PICASSO.  In these bubble chamber experiments, operation at varying temperature or pressure essentially integrates the recoil spectrum above some threshold.  The background from alpha decays is known to be a flat spectrum above some specific temperature or pressure and is fit to in the data.  For the broadest MDDM spectra (see Fig.~\ref{fig:Gespectra}),  the dark matter signal looks similar to this alpha background. Unfortunately, this complicates background subtraction and reduces the present sensitivity to these models.

Intriguingly, momentum dependent scattering can also modify the interpretation of dark matter explanations of DAMA's annual modulation signal and exclusions from other direct detection experiments.   The annual modulation signal \cite{Drukier:1986tm,Freese:1987wu}, originally seen at the DAMA experiment \cite{DAMA}, has recently been confirmed by DAMA/LIBRA \cite{Bernabei:2008yi}. On the other hand, limits from XENON and CDMS  strongly constrain the simplest dark matter interpretation of the DAMA experiment: a signal resulting from the SI scattering of a WIMP. Explanation of the DAMA signal with spin-dependent scatterings \cite{FreeseOld,Savage:2008er} is now also strongly constrained by COUPP and PICASSO.
 
Fig.~\ref{fig:SIq2} shows that adding momentum dependence to the SI interactions can only weaken, but not eliminate the limits other experiments put on DAMA explanations at the 90\% confidence level, at least within a Maxwellian halo model. Employing the caveats discussed in \cite{Chang:2008xa}, alternative statistical techniques \cite{Savage:2008er} or a non-Maxwellian halo \cite{Fairbairn:2008gz} might allow a window at low mass when combined with these new effects.

In light of this, we now focus discussion on the scenario with the weakest direct detection limits, SD-proton scattering, shown in Fig.~\ref{fig:SDp}.  These plots show that the relative importance of different experiments can invert as one adds momentum dependence, for precisely the reasons described above.  In fact, the normal SD-proton case \cite{FreeseOld,Savage:2008er} which is ruled out by PICASSO is allowed for the $q^4$ scenario.  Interestingly, these factors are also able to improve the fit with DAMA's spectral shape, so that there are new masses that can now fit the DAMA spectrum.  In particular, the mass region at $\sim 40-60\; \gev$ would have normally had a shape that was inconsistent with DAMA.   Since these momentum factors suppress the low energy scattering, the constraint from DAMA's unmodulated event rate \cite{Chang:2008xa} is also weakened, leading to better consistency with DAMA's full data set. For these plots, we assumed a mediator mass of 1 GeV and 100 MeV.  As we will discuss later in the next section, a  lighter mediator mass of $O(100)$ MeV is more suited to generate cross sections of this size.  As seen in Fig.~\ref{fig:SDp}(c), for this lighter choice of mass, the 10 GeV DM mass region survives, but the KIMS limit cuts into about half of the higher mass region. We should note that our approach to the KIMS limits is not aggressive, and does not yield as strong a limit as that in \cite{KIMS}. Consequently, a more aggressive limit might also be able to exclude this region as well. For mediator masses much less than an MeV,  the momentum independent case is recovered since the $q^2$ factors cancel in Eqn.~(\ref{eqn:newdR}), at least in the range that the first Born approximation is valid.

One important point about these scenarios is that the expected relationship between direct and indirect detection signals breaks down. Typically, one assumes that the annihilation proceeds into some Standard Model final state. Here, since we rely upon the light mediator for our interaction, it provides an annihilation channel. Then, if the mediator is $\lesssim \gev$ in mass, it is natural for it to decay dominantly to e.g., electrons, muons and pions.  The limits from Super--Kamiokande WIMP capture are then trivially evaded \cite{Itay,Menon:2009qj}.  

\section{Model Building Constraints}
For the dark matter scattering to display the novel phenomenology discussed here, the scattering rate must be dominated by the new operators, and not the typical SI or SD coupling.  This is not a trivial requirement.  For comparable coefficients, the scattering mediated by operators of Eqns.~(\ref{eqn:operators1})--(\ref{eqn:operators}) are suppressed by powers of the velocity or momenta relative to Eqns.~(\ref{eqn:SI}) and (\ref{eqn:SD}).  It is possible, however, that the coefficients of these new operators are much larger than the coefficients of the other operators.    We discuss this further below.

 If the signal is to be observable at near-future experiments, the $q^{2n}$ suppression must be compensated by a large coefficient for the operator.  This could be due in part to particularly large couplings of a mediator to the Dark Sector or a local over-density of the dark matter, but the simplest way to get an enhancement is just for the mediator mass to be small:  $dR/dE_{R}  \propto m_{\phi}^{-4}$.   The necessary $m_{\phi}$ depends on the amount of $q^2$ suppression. If there is a single $q^2$, as in \op{1} and \op{2}, a mediator mass of a $m_\phi \sim {\rm few} \;\gev$, $O(1)$ couplings to the DM while having Yukawa suppression on quark side, one finds a $\sim 10^{-36} {\rm \; cm^2}$ cross section.  This is near the interesting region for $\op{2}$.  In the case of coherent scattering off of nuclei  (as for \op{1}), this would actually already be strongly ruled out by CDMS and XENON for a large mass range, a viable $10^{-44} {\rm \; cm^2}$ cross section would require something like $m_\phi \sim 100 \;\gev$.   For $q^4$ suppression, the mass has to be $O(100) \;\Mev$ to get a $\sim 10^{-36} {\rm \; cm^2}$ cross section.
 
In specific cases, large contributions  to $\op{1}$ - $\op{4}$ can be expected.  If there is a light pseudoscalar present that couples both to dark matter and to quarks then $\op{1}$ - $\op{3}$ can be generated through its exchange without generation of either $\op{SI}$ or $\op{SD}$.  If there is no parity violation, then the expectation is that $\op{3}$ dominates.  On the other hand, if parity violation is present, then it is plausible that the coefficients of $\op{1}$ - $\op{3}$ could all be comparable.  In this case, it is likely that it would be easiest to probe $\op{1}$ because of its coherent scattering off of nuclei.  Alternatively, parity violation might be confined to couplings in the dark matter sector.  In this case, pseudoscalar exchange could dominantly induce $\op{2}$.   We note that if the  light pseudoscalar is naively realized as a pseudo-Goldstone, it is difficult to sufficiently suppress the contributions to \op{SI}.  Contributions are induced by exchanging the scalar whose vacuum expectation $f_\phi$ value breaks the global symmetry and made the $\phi$ light. 
 
Interactions with only $q^2$ suppression can conceivably still dominate over standard interactions without significant model-building efforts. The simplest example comes from charge-radius or dipole couplings to a composite WIMP, whose constituents are charged under a new, dark gauge group \cite{cidm}. This generates the phenomenology of \op{1} straightforwardly, although typically with a coupling to $Z^2$ instead of $A^2$. If the mediation arises through a PGB, \op{1,2} can dominate over the scalar exchange as only one vertex will be suppressed by $f_\phi$, while the scalar exchange is suppressed by $f_\phi^2$. 

The most challenging model-building comes in realizing the $q^4$ suppressed interaction, without inducing SI scattering from the accompanying scalar mediator. While this seems difficult from the perspective of a standard PGB, it can arise fairly simply in SUSY theories. While PGBs are a natural way to realize a shift symmetry, such a shift symmetry could simply be present in the theory from other origins. For instance, in theories with $N=2$ SUSY in the gauge sector (e.g., \cite{Fox:2002bu}), there is a chiral superfield partner for every gauge boson. The pseudoscalar contained in it possesses a shift symmetry which can be thought of as  a higher-dimensional gauge symmetry, compactified on an $S_1/Z_2$ orbifold. SUSY breaking will make the associated scalar massive.  This can arise either from $F-$term breaking, through the operator $X^\dagger X (\phi + \phi^\dagger)^2$ (with $X$ a spurion that gets the non-zero $F$-term, and $\phi$ a superfield containing the PGB), or from $D$-terms, through $W^\alpha W'_\alpha \phi$ (with $W^\alpha$ the $U(1)_{Y}$ supersymmetric  field strength, and  $W'_\alpha$ the supersymmetric field strength that gets a non-zero $D$-term).   Even in the presence of this supersymmetry breaking, the pseudoscalar remains massless, and will only pick up a mass radiatively through diagrams violating {\em both} the shift symmetry and SUSY. Thus, the scalar contribution can be effectively decoupled from the strength of the pseudoscalar-mediated $q^4$ interactions.
  
Finally, a sizable coefficient for $\op{4}$ can be generated in theories with a light gauge boson that couples to the dark matter and mixes with the $B^{\mu}$ gauge field of the Standard Model. 

There are model-dependent constraints on light pseudoscalar mediators.  In particular,  searches for axions can apply.  For particles in the GeV range, the process $\Upsilon \rightarrow \gamma \phi$ is relevant.  These branching ratios are constrained to be in the range  $10^{-5} - 10^{-6}$; the precise bound depends on the final state of $\phi$ decay, $\phi \to \mu\bar{\mu},\tau\bar{\tau}$ or invisible, \cite{BaBarUpsilonMu,BaBarUpsilonTau,CLEOUpsilonMiss}.  If the $\phi$ has couplings comparable to Standard Model Yukawas,  the branching ratio is a $few \times 10^{-5}$ for masses well below the $\Upsilon$ mass.  Thus, these bounds constrain the $\phi$ coupling to $b$ quarks to be somewhat smaller than the Standard Model Yukawa coupling. For lighter mediators, depending on the flavor structure of the $\phi$ couplings, $K \rightarrow \pi \pi \phi$ may be relevant.  The rate for the potentially more stringent process $K^{+} \rightarrow \pi^{+} \phi$ is suppressed -- a pure pseudoscalar coupling does not mediate this process, see for e.g., \cite{Deshpande:2005mb}.  The dominant contribution to this decay comes from $\pi - \phi$ mixing \cite{Weinberg:1977ma}, which is model dependent.  In cases where the pseudoscalar couples only to 3rd generation quarks, the Kaon decays are absent; however, 
the Upsilon constraints still apply.  Following the procedure in \cite{Cheng:1988im}, we find that 3rd generation couplings alone can generate a detectable rate, as pseudoscalar couplings to heavy quarks generate a coupling to $G\widetilde{G}$ \cite{Anselm:1985cf}.  Incidentally, in general, experimental uncertainties (in particular, the light quark contribution to the proton spin $\Delta \Sigma$) and parameters like $\tan \beta$ allow a proton dominated coupling to be generated.  In the minimal case of 3rd generation couplings, the $\phi$ decays to two photons with a decay length $c\tau \sim 1 {\rm \;m}$.   While couplings to leptons are not necessary to implement the scenario at hand,  if the mediator couples with Yukawa strength to the muon, requiring that the magnitude of the  contribution to the muon $g-2$ is  no larger than the current discrepancy between theory and experiment $ |\delta a_{\mu}|  <  290 \times 10^{-11}$   enforces $m_{\phi} >  300$ MeV \cite{Deshpande:2005mb}.   Finally with couplings to electrons, it is possible to search for $e^+ e^- \to \phi \gamma$ for either invisible or electron decays of $\phi$ \cite{Borodatchenkova:2005ct}.  However, for pseudoscalar $\phi$, suppression by the electron yukawa coupling makes the production rate below the projected sensitivities \cite{Borodatchenkova:2005ct}.

 \section{Conclusions}
As the sensitivity of new dark matter direct detection experiments continues to increase at a rapid pace, the ability to test for new scenarios for dark matter will grow simultaneously. Present experiments are optimized to search for WIMPs with signals that peak at low nuclear recoil energies. In contrast, models with momentum-suppressed interactions (MDDM), have spectra that peak at intermediate energies, thus changing the expected signals and relative strengths of various direct detection experiments. While interesting scenarios, specifically inelastic dark matter, have been proposed with spectra that peak at high recoil energy, we find that the scenarios with this feature are more ubiquitous than previously thought. In models with new light vectors or pseudoscalars, momentum dependent interactions can be large, and this phenomenology can be present.

A simple parameterization captures much of the relevant phenomenology. Specifically, one can replace $dR_i/dE_R$ with $(q_{100}^2)^n dR_i/dE_R$, where $i$ indexes the interaction type and $q_{100}$ is the momentum transfer in units of 100 MeV. While additional features can arise at low mediator masses, this parameterization is sufficient to reproduce the peaking in the spectrum, and provides a convenient way to compare different experiments. In analyzing the presently allowed parameter space in this way, we find that momentum dependent couplings can open allowed ranges of parameters for DAMA with dominantly spin-dependent proton couplings, if accompanied by an additional $q^4$ suppression.

Whatever model of dark matter nature has chosen to realize, it is important to be cognizant of the wide range of possible phenomenology, so that possible signals are not missed or attributed to backgrounds. The framework of MDDM provides motivation, and a prescription to study and constrain these models in the future.

\vskip 0.15in
{\noindent Note added:}  As this paper was being finished, we became aware of \cite{AmiFF}, which appeared in the arXiv and discusses momentum dependent interactions arising from the couplings to one or more new gauge bosons, and their ability to explain DAMA from spin-independent interactions.

\begin{acknowledgments}
\section{Acknowledgements} 
\noindent We would like to thank Chris Savage for useful discussions, and for providing us with code implementing the spin-dependent form factors. We thank Peter Cooper from COUPP and Viktor Zacek and Sujeewa Kumaratunga from PICASSO for information on their analyses.  The work of SC is supported under DOE Grant \#DE-FG02-91ER40674.  The work of AP is supported under DOE Grant \#DE-FG02-95ER40899 and by NSF CAREER Grant NSF-PHY-0743315.  The work of NW is supported by NSF CAREER grant PHY-0449818 and DOE OJI grant \#DE-FG02-06ER41417.
 \end{acknowledgments}

\bibliography{MDDM}
\bibliographystyle{apsrev}
\end{document}